\documentclass[11pt,twoside]{article}
\usepackage{asp2010}

\resetcounters

\newcommand{\be}{\begin{equation}}

\newcommand{\ee}{\end{equation}}

\newcommand{\lavec}[1]{{\mathbf{#1}}}         

\def\al{Alfv\'en\ }

   \def\b{{\lavec{b}}} \def\u{{\lavec{u}}}     

\begin{document}

\title{High Lundquist Number Resistive MHD Simulations of Magnetic Reconnection: Searching for Secondary Island Formation}
\author{C. S. Ng and S. Ragunathan
\affil{Geophysical Institute, University of Alaska Fairbanks, Fairbanks, AK 99775, USA}
}

\begin{abstract}
Recently, secondary island formation due to the tearing instability of the Sweet-Parker current sheet 
  was identified as a possible mechanism that can lead to fast reconnection 
  (less sensitive dependence on Lundquist number $S$) both in numerical simulations using 
  Particle-in-Cell (PIC) method [Daughton et al. 2009], as well as using resistive 
  magnetohydrodynamics (MHD) [Lapenta 2008; Bhattacharjee et al. 2009]. 
This instability is thought to appear when $S$ is greater than a certain threshold. 
These recent results prompt us to perform more resistive MHD simulations of a basic 
  reconnection configuration based on the island coalescence instability, 
  using much higher resolutions and larger $S$. 
Our simulations are based on a fairly standard pseudo spectral code, 
  which has been tested for accuracy, convergence, and compared 
  well with codes using other methods [Ng et al. 2008]. 
In our simulations, formation of plasmoids were not found, 
  except when insufficient resolution was used,
  or when a small amount of noise was added externally.
The reconnection rate is found to follow the Sweet-Parker scaling when no noise is added,
  but increases to a level independent of $S$ with noise, 
  when plasmoids form.
Latest results with $S$ up to $2 \times 10^5$ will be presented. 
\end{abstract}

\section{Introduction}   
\label{Intro}
It is well known within the space physics community that the classic Sweet-Parker theory of magnetic reconnection 
  predicts a reconnection rate that is too slow for most energetic phenomena, such as solar flares or magnetospheric
  storms.
Strictly speaking, the Sweet-Parker theory is based on steady state two dimensional (2D) incompressible 
  resistive magnetohydrodynamics (MHD), with a uniform classical resistivity $\eta$, 
  although it might also apply approximately in a more general system.
It predicts a reconnection rate that is proportional to $S^{-1/2}$, where $S = \mu_0 V_A L /\eta$ is the Lundquist number,
  with $V_A$ being the \al speed based on the upstream magnetic field strength, 
  and $L$ being the typical length scale of system.
Obviously such reconnection rate would be too small for large $S$, which is usually the case since $L$ is very large for 
  most systems in space or astrophysics, and $\eta$ is usually small for high temperature plasmas. 
Such a small reconnection rate is related to the formation of a long (to the order of the system size) but 
  very thin (with thickness $\delta$ also proportional to $S^{-1/2}$) reconnection current sheet,
  and so the inflow velocity is restricted due to the constrain of the continuity of mass. 
It is also well known that there have been many different theories or simulations 
  over many years that can have a much higher reconnection rate, 
  based on relaxing one or more restrictions in the Sweet-Parker model, e.g., effects due to slow shock 
  in compressible MHD, anomalous or non-uniform resistivity, Hall or two-fluid MHD, turbulence, 
  and kinetic physics using direct particle-in-cell (PIC)
  simulations (too many to cite in this short paper).   
In the last few years however, there has been a renewal of interest in reconnection in the Sweet-Parker setting,
  since a much higher reconnection rate (apparently independent of $S$) 
  has been reported in some studies \citep{2008PhRvL.100w5001L, 2009MNRAS.399L.146L, 2009PhPl...16k2102B, 
  2009PhPl...16l0702C, 2010PhPl...17f2104H}.
A major reason for such an increase in reconnection rate is considered to be due to the tearing instability of the Sweet-Parker
  current sheet in the high-Lundquist number regime, and the formation and ejection of secondary islands (plasmoids), 
  which has also been observed recently in PIC simulations \citep{2009PhRvL.103f5004D},
  but actually has been observed for many years, e.g., \citep{1986JGR....91.6807L}.
Incidentally, one of the authors of this paper (Ng) has seen secondary island formation in reconnection simulations
  since 1996, but always regarded that to be due to insufficient resolution and thus such an observation was not reported.
In this study, we plan to revisit this problem by performing careful simulations using as high resolution as we can (up to $8192^2$),
  for $S$ up to $2 \times 10^5$.
The simulation geometry is as described in \citep{2008ApJS..177..613N} and is essentially 
  the same as what was used in \citep{2009PhPl...16k2102B}. 
What we found is that we have not been able to see the tearing instability of the Sweet-Parker current sheet, 
  as well as the formation of plasmoids, despite the $S$ value used is in the range that these are 
  seen in \citep{2009PhPl...16k2102B}. 
The reconnection rate found over a range of $S$ follows the Sweet-Parker scaling of $S^{-1/2}$.

Moreover, in order to make sense of why we obtain different results, we performed another set of simulations
  with the same settings, except, random noise was now added in small scales.
With that, plasmoids do appear and the reconnection rate found is roughly independent of $S$.
This idea of reconnection with noise is similar to the concept of turbulent reconnection, 
  which has been around for a while \citep{1986PhFl...29.2513M, 1999ApJ...517..700L}.
There have also been a number of new studies on turbulent reconnection, e.g., \citep{2009ApJ...700...63K}.
Note that some studies mentioned above also have noise added \citep{2009MNRAS.399L.146L, 
  2010PhPl...17f2104H}, although there is still no universal understanding on the importance of the added noise.

In Section~\ref{model}, we describe briefly our simulation method and geometry. 
Our numerical results will be presented in Section~\ref{results}, 
  with a discussion in the final Section.

\section{Simulation Model}   
\label{model}
Our results are based on numerical simulations of the 2D incompressible MHD equations, 
which can be written in the following normalized form,
\begin{equation}
\partial_t {\Omega} + [\phi,\Omega]  = [A,J]
    + \nu \nabla_{\perp}^2 {\Omega} , \hspace{0.3in}
\partial_t {A} + [\phi,A]  = \eta \nabla_{\perp}^2 {A} ,
\label{eq_induction-2D} \end{equation}
where $A$ is the flux function so that the magnetic field is expressed as $\b = \lavec{\hat{z}} +  {\nabla}_{\perp}A \times \lavec{\hat{z}}$,
$\phi$ is the stream function so that the velocity field is $\u =  {\nabla}_{\perp}\phi \times \lavec{\hat{z}}$, 
$\Omega= -\nabla^2_{\perp}\phi$ is the  $z$-component of the vorticity, 
$J = -\nabla^2_{\perp}A$ is the  $z$-component of the current density,  
and $[\phi,A]  = \partial_y\phi\partial_xA - \partial_yA\partial_x\phi$. 
In this normalized form, the resistivity $\eta$ is essentially the inverse of the Lundquist number.
A more precise conversion of $S = 2\pi \bar{A} /\eta$ should be used when comparing with other studies, 
  where $\bar{A} = 0.4$ is the maximum magnitude of $A$
  used in the initial condition $A(x,y,t=0) = \bar{A} \sin(2\pi x) \sin(2\pi y)$.
In all our runs, we have chosen viscosity $\nu$ to be the same as $\eta$.
Periodic boundary conditions are used in both the $x$- and $y$- directions,  
  with $0 \leq x \leq 1$ and $0 \leq y \leq 1$.
A small flow is added initially to start the growth of the island coalescence instability. 

The simulation code is based on the pseudo spectral method using 
   fast Fourier transform (FFT) on a 2D bi-periodic domain. 
It is de-aliased by the standard 2/3 rule.  
The nonlinear term is calculated in the physical space on a uniform grid of collocation points.
A second order predictor-corrector method is used for time integration. 
The code is parallelized using a parallel version of the FFT.  
This code has been tested for convergence, with accuracy of results compared 
  with two other methods \citep{2008ApJS..177..613N} and found to be
  highly accurate in tests of conservation laws (with fractional errors usually three to four orders of 
  magnitude smaller than those using a finite difference method), 
  as expected from a spectral code.
However, the price to pay for a more accurate code is that it requires much longer CPU time to run,
  especially in high resolution cases due to the use of a uniform grid, 
  as well as explicit time integration. 

\begin{figure}[!ht]
\plottwo{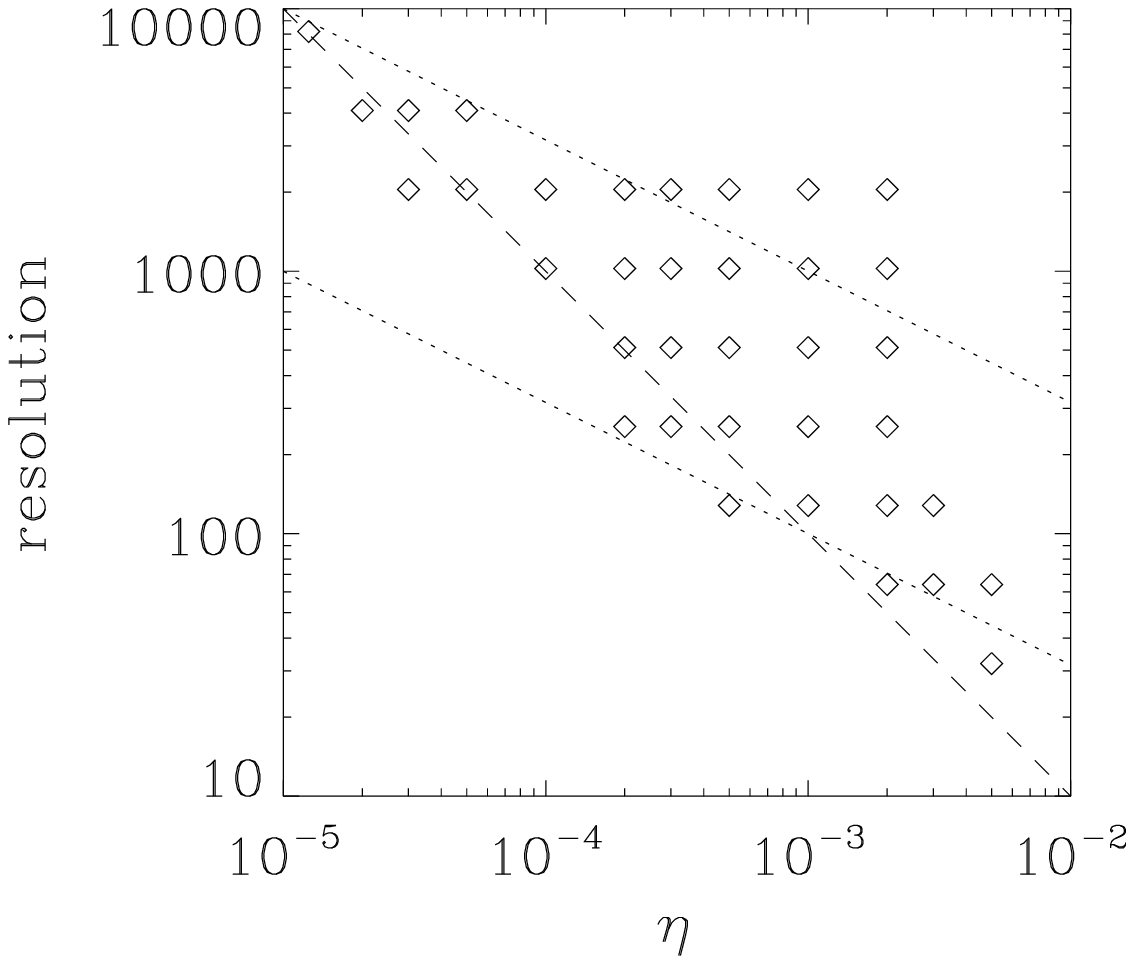}{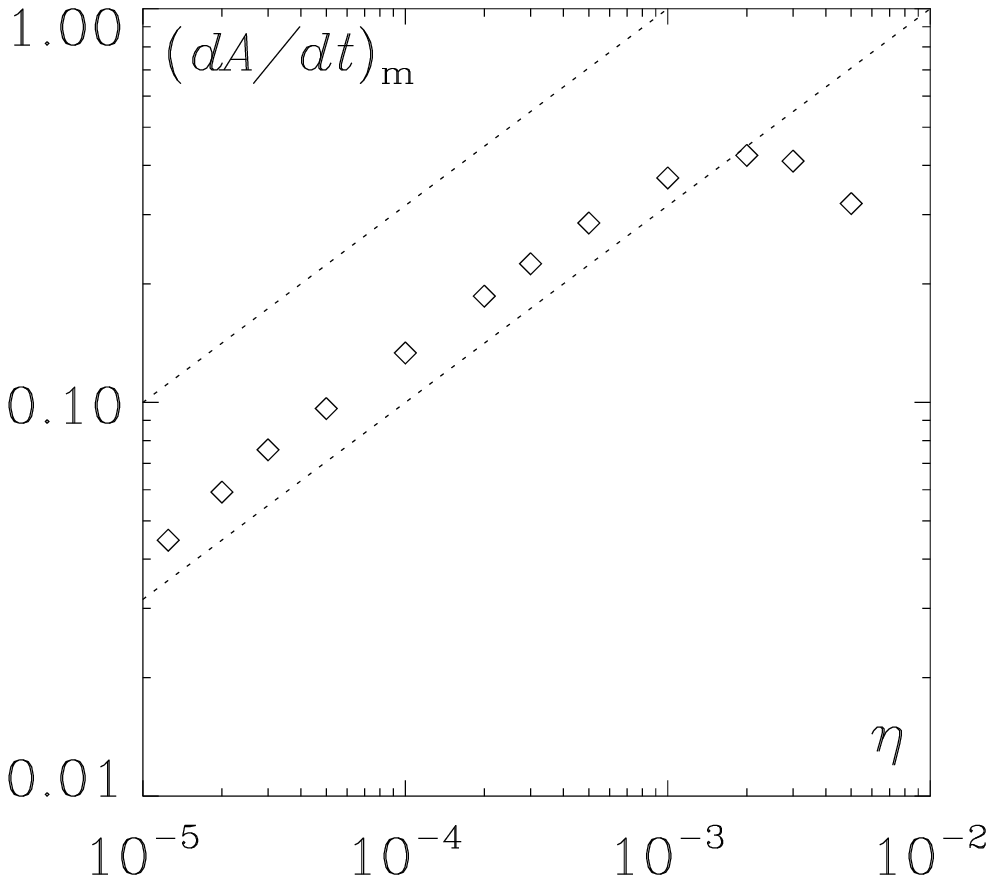}
\label{resol}
\caption{(a) Left: linear resolution of adequately resolved runs for different resistivity $\eta$. 
Dotted lines that scale with $\eta^{-1/2}$ and a dashed line that scales with $\eta^{-1}$ are plotted for comparison. 
(b) Right: maximum reconnection rate over the whole duration of simulation for cases with different $\eta$.
Dotted lines scales with $\eta^{1/2}$.}
\end{figure}

\section{Simulation Results}   
\label{results}

We have run many cases for different $\eta$ and linear resolution (grid is at resolution square).
Fig.~1(a) shows all cases (indicated with diamonds) that we have run and were adequately resolved.
Note that for $\eta$ at larger values, different resolutions have been used 
  up to $2048^2$ or $4096^2$ so as to test for convergence. 
It is confirmed that as long as a run is adequately resolved, the numerical solution only
  change slightly as resolution increases. 
Moreover, we have also run cases with resolution lower by a factor of two of the lowest
  resolution cases shown in the figures.
It usually was quite obvious that those runs were not well-resolved. 
Some of them indeed have plasmoids, which disappear as resolution is increased. 
Another remark on Fig.~1(a) 
  is that the linear resolution seems to change from
  scaling with $\eta^{-1/2}$ (as shown in dotted lines) in the large $\eta$ regime to 
  scaling with $\eta^{-1}$ (as shown in the dashed line) in
  the small $\eta$ regime, with the transition at around $\eta \sim 10^{-4}$.
More careful examination of simulation outputs indicates that this is because at 
  smaller $\eta$, the length scale (proportional to $\eta$) of other structures
  becomes smaller than the width of the Sweet-Parker current sheet, 
  which scales with $\eta^{1/2}$.
Such small scale structures are found to be at the locations where the reconnection
  outflows collide with stationary structures.
Moreover, these small scale structures have orientation perpendicular to the 
  Sweet-Parker current sheet. 
Therefore, resolving the Sweet-Parker current sheet itself
  (say by using nonuniform grids with denser grid lines along the Sweet-Parker
  current sheet) is not enough for resolving the whole simulation domain.
Obviously such a change of dependence of resolution on $\eta$ means it is 
  even more difficult to simulate high-Lundquist number cases,
  and one needs to pay attention to such a resolution demand when simulating
  these cases. 
  
In Fig.~1(b), 
  we plot the maximum reconnection rate of each of these runs over the 
  whole duration of simulation.
Since we are interested in the high Lundquist number regime, 
  we measure the reconnection rate simply by looking at the value of $A$ at the central X-point, $A_0$,
  so that reconnection rate is given by $dA_0/dt$.
From this plot, it is clear that the data follow a Sweet-Parker scaling of $\eta^{1/2}$, 
  as indicated by the dotted lines.
This means that we have not found increased reconnection rate even for the case with the smallest 
  resistivity $\eta = 1.25 \times 10^{-5}$, or $S \sim 2 \times 10^5$.
More careful examination of the outputs, by plotting contours of $A$ and $J$ in movies,
  shows that there were no plasmoids forming in these runs.
Fig.~2(a) 
  shows $A_0$ as functions of time $t$ for cases with
  different $S$ showing explicitly how $A_0$ grows slower with $t$ as $S$ gets larger.

\begin{figure}[!ht]
\plottwo{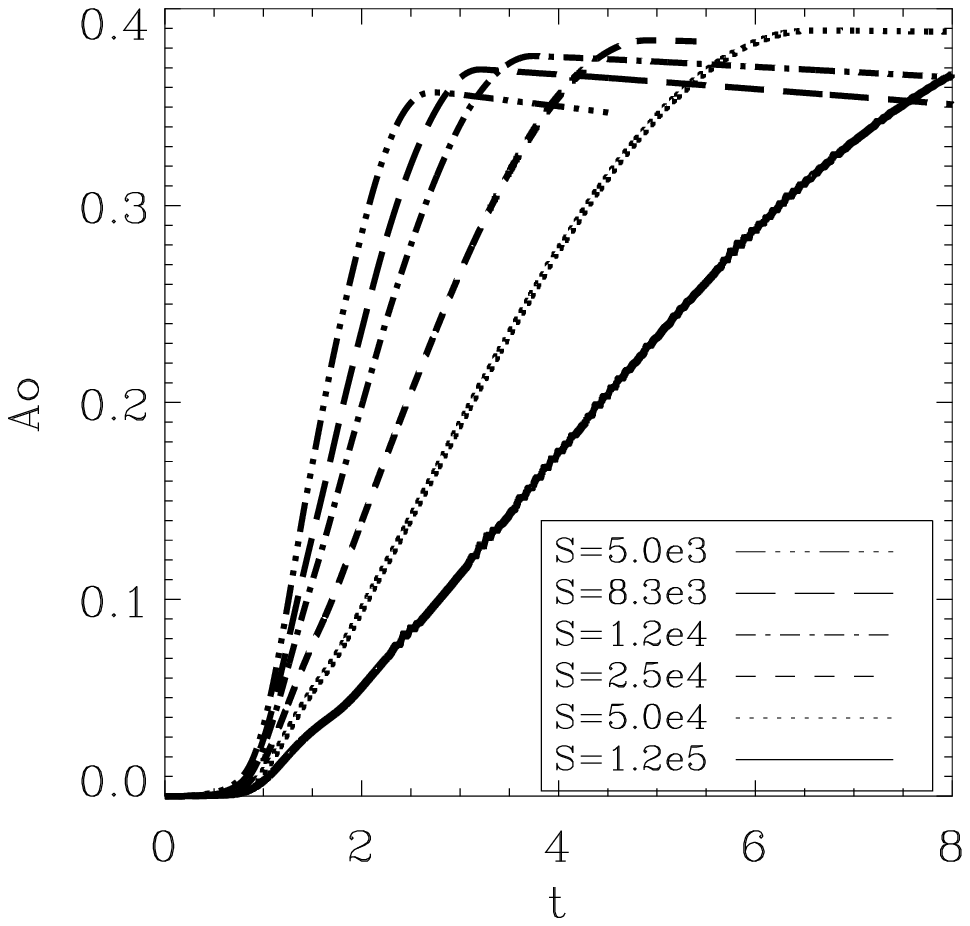}{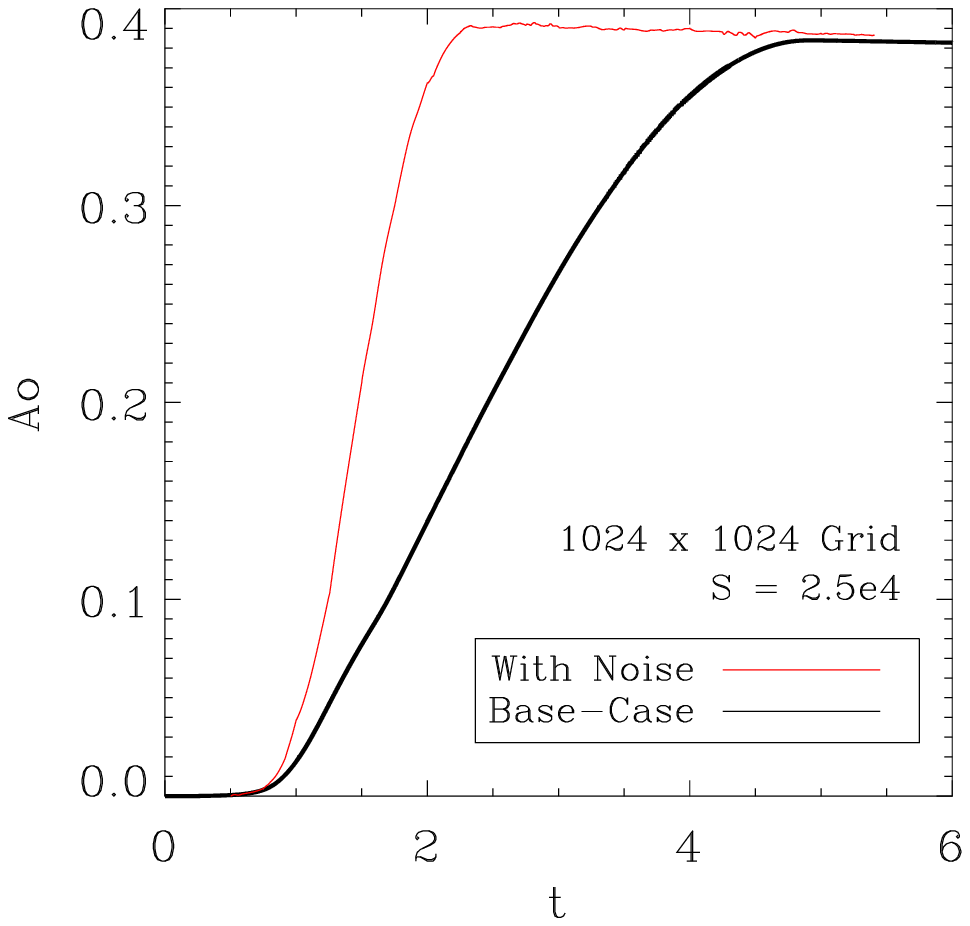}
\label{A0comp}
\caption{(a) Left: $A_0$ as functions of $t$ for different $S$.
(b) Right: $A_0$ as functions of $t$ for the $S = 2.5 \times 10^4$ case using 
$1024^2$ resolution with noise added (red) and without noise (black).  }
\end{figure}

Since no plasmoids were observed with the well-resolved runs, 
  and that our experience shows they can form in under-resolved cases, 
  it is plausible that the inherent noise present in other numerical methods could 
  have led to the formation of plasmoids in those simulations. 
To explore this possibility further, we have run our cases again with external 
  random noise added throughout the whole simulations. 
The method of adding noise is similar to the procedure outlined in \citep{1999PhFl...11.1880A, 2009ApJ...700...63K}.  
Noise is added to the simulation in spectral space with a characteristic wavenumber 
  (depending on the system parameters) and an amplitude that ensures that the noise 
  added is always of a small fraction of the kinetic and magnetic energies at any time.  

\begin{figure}[!ht]
\plotone{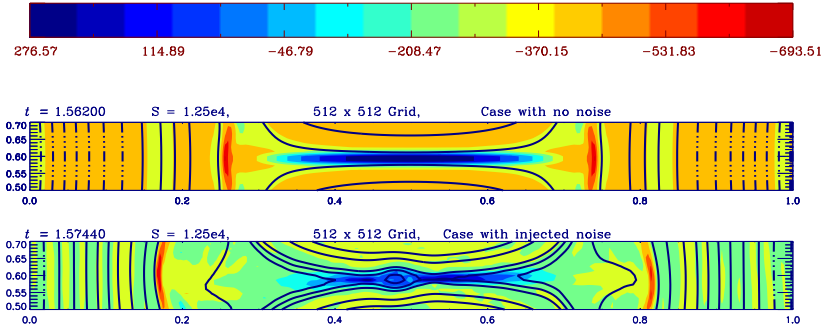}
\label{plasmoid-comp}
\caption{(a) Top: Contour lines of $A$ for the case $S = 1.25 \times 10^4$ with no noise.
Color filled contours are for $J$ to show the location of the current sheet (dark blue region).
(b) Bottom: plot for the same case with noise added when a plasmoid is formed. The two plots are at roughly the same time.
 }
\end{figure}

Fig.~2(b) 
  shows an example of this study with $A_0$ vs $t$ plotted for the $S = 2.5 \times 10^4$ case using 
  $1024^2$ resolution with noise added (red) and without noise (black).
Obviously one can see that the reconnection rate is enhanced in the case with noise added.
An examination of outputs shows plasmoid formation and ejection, 
  e.g. as shown in Fig.~\ref{plasmoid-comp}, so that instantaneous reconnection rate 
  fluctuates a lot at the same time, but averaged out to be a higher rate.
  
\begin{figure}[!ht]
\plottwo{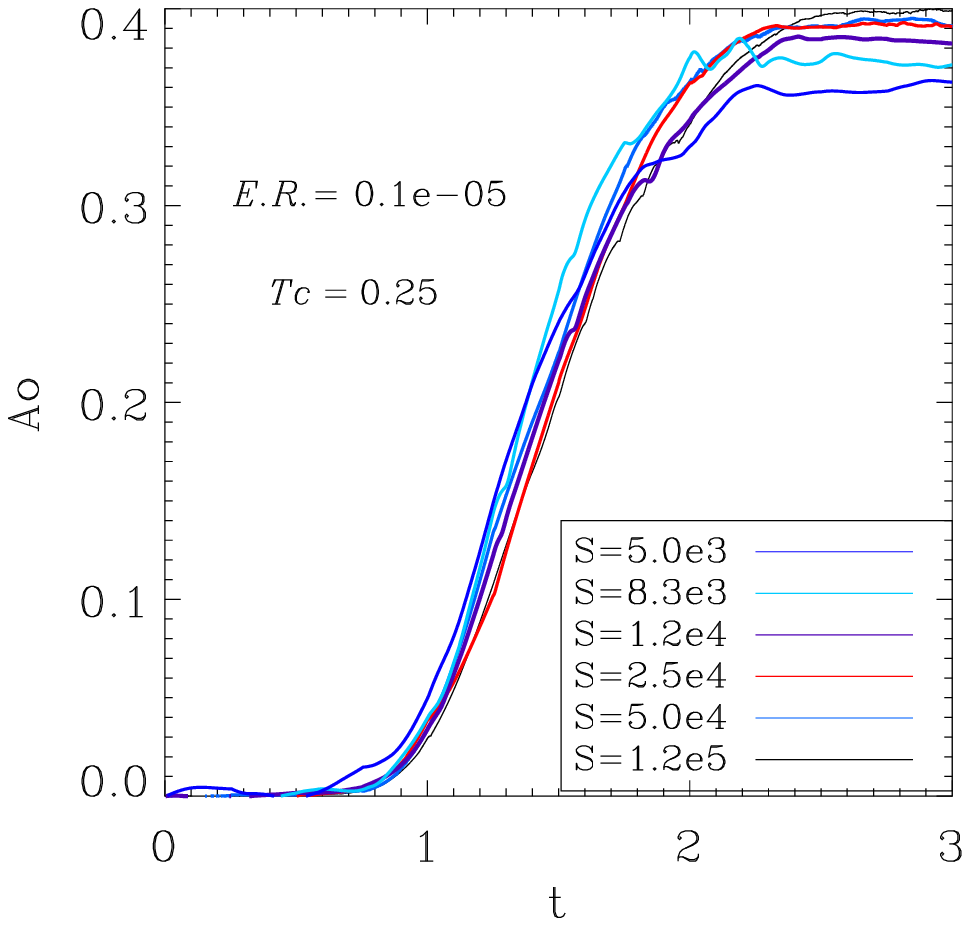}{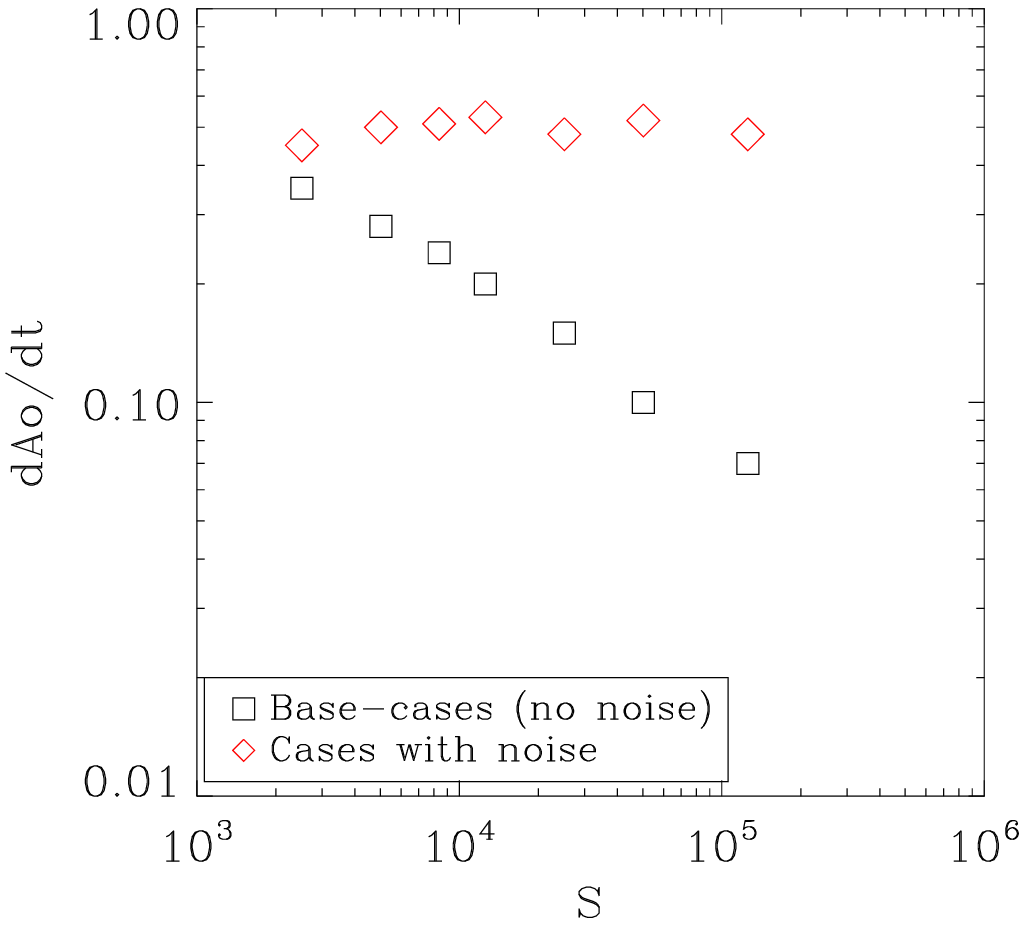}
\label{A0noise}
\caption{(a) Left: $A_0$ as functions of $t$ for different $S$ with noise added.
(b) Right: average reconnection rate $dA_0/dt$ for different $S$ with noise added 
(red diamonds) and without noise (black squares). }
\end{figure}

In Fig.~4(a), 
  $A_0$ as functions of $t$ for more cases with different $S$ 
  are plotted on the same graph, showing that the reconnection rate does not 
  change much even as $S$ changes a lot.
These cases are simulated using different resolutions, and thus the scale at which 
  noise is added also changes.
However, to have a more consistent comparison, such noise is kept to have energy 
  roughly of the same ratio to the total energy ($\sim 10^{-6}$), 
  and that the correlation time of the random noise is kept at about 0.25 for all cases.
To characterize reconnection rates for all these cases, 
  the averaged values of $dA_0/dt$ over the connecting phase are plotted on 
  Fig.~4(b) 
  in red diamonds, 
  together with reconnection rate for the original cases without added noise (black squares).
This shows clearly that the cases with noise have a reconnection rate roughly   
  independent of $S$, and are at a level much larger than the reconnection rate without 
  noise, especially in the large $S$ limit,
  since the reconnection rate is consistent with a Sweet-Parker scaling of $S^{-1/2}$.

\section{Discussion and Conclusion}   
\label{discuss}

We have presented our simulation results showing that reconnection rate in resistive MHD
  do follow the Sweet-Parker scaling over a significant range of Lundquist number up to 
  $S \sim 2 \times 10^5$, if runs are well-resolved, and that no external noise is added.
In these runs, apparently no tearing instability occurs at the Sweet-Parker current sheet,
  and thus no plasmoid formation is observed.
In trying to understand why our results differ from results from other studies, 
  we repeated our runs with random noise added externally.
We have shown that plasmoids do form in these cases, 
  resulting in a reconnection rate roughly independent of $S$,
  and thus much higher than the case with no noise in the large $S$ limit.
While these results do show that plasmoid formation in reconnection of 
  this setting occurs only when external noise is added,
  we do not claim that this also explains what were observed in simulations of other studies.
This result does highlight the importance of making sure that a simulation is run 
  with sufficient resolution, e.g., by checking convergence with higher resolutions.
  
We remark that we have not done a full study of how reconnection rate depends on
  properties of the added noise, which is in fact not an easy task.
For example, one surely can expect that the effect of noise would be negligible if its total
  energy tends to zero.
We have confirmed this by running cases at different levels of noise energy,
  but we have not run enough cases so far to show a dependence of reconnection
  rate on the noise energy.
More importantly, the next step we need to investigate is what physical effects are
  responsible for generating such noise that can enhance reconnection rate.
Can we stay within MHD, e.g., using turbulence, or do we need some external physics?

\acknowledgements The authors thank Drs. Y.  Huang and A. Bhattacharjee for valuable discussion. This work is supported by a NASA grant NNX08BA71G, and a NSF grant AGS-0962477. This work was supported in part by a grant of HPC resources from the Arctic Region Supercomputing Center and the University of Alaska Fairbanks.

\bibliography{Ng-astronum2010}

\begin{thebibliography}{}
\expandafter\ifx\csname natexlab\endcsname\relax\def\natexlab#1{#1}\fi
\expandafter\ifx\csname url\endcsname\relax
  \def\url#1{\texttt{#1}}\fi
\expandafter\ifx\csname urlprefix\endcsname\relax\def\urlprefix{URL }\fi
\providecommand{\eprint}[2][]{\url{#2}}

\bibitem[{{Alvelius}(1999)}]{1999PhFl...11.1880A}
{Alvelius}, K. 1999, Phys. Fluids, 11, 1880

\bibitem[{{Bhattacharjee} et~al.(2009){Bhattacharjee}, {Huang}, {Yang}, \&
  {Rogers}}]{2009PhPl...16k2102B}
{Bhattacharjee}, A., {Huang}, Y., {Yang}, H., \& {Rogers}, B. 2009, Phys.
  Plasmas, 16, 112102

\bibitem[{{Cassak} et~al.(2009){Cassak}, {Shay}, \&
  {Drake}}]{2009PhPl...16l0702C}
{Cassak}, P.~A., {Shay}, M.~A., \& {Drake}, J.~F. 2009, Phys. Plasmas, 16,
  120702

\bibitem[{{Daughton} et~al.(2009){Daughton}, {Roytershteyn}, {Albright},
  {Karimabadi}, {Yin}, \& {Bowers}}]{2009PhRvL.103f5004D}
{Daughton}, W., {Roytershteyn}, V., {Albright}, B.~J., {Karimabadi}, H., {Yin},
  L., \& {Bowers}, K.~J. 2009, Phys. Rev. Lett., 103, 065004

\bibitem[{{Huang} \& {Bhattacharjee}(2010)}]{2010PhPl...17f2104H}
{Huang}, Y., \& {Bhattacharjee}, A. 2010, Phys. Plasmas, 17, 062104

\bibitem[{{Kowal} et~al.(2009){Kowal}, {Lazarian}, {Vishniac}, \&
  {Otmianowska-Mazur}}]{2009ApJ...700...63K}
{Kowal}, G., {Lazarian}, A., {Vishniac}, E.~T., \& {Otmianowska-Mazur}, K.
  2009, \apj, 700, 63

\bibitem[{{Lapenta}(2008)}]{2008PhRvL.100w5001L}
{Lapenta}, G. 2008, Phys. Rev. Lett., 100, 235001

\bibitem[{{Lazarian} \& {Vishniac}(1999)}]{1999ApJ...517..700L}
{Lazarian}, A., \& {Vishniac}, E.~T. 1999, \apj, 517, 700

\bibitem[{{Lee} \& {Fu}(1986)}]{1986JGR....91.6807L}
{Lee}, L.~C., \& {Fu}, Z.~F. 1986, \jgr, 91, 6807

\bibitem[{{Loureiro} et~al.(2009){Loureiro}, {Uzdensky}, {Schekochihin},
  {Cowley}, \& {Yousef}}]{2009MNRAS.399L.146L}
{Loureiro}, N.~F., {Uzdensky}, D.~A., {Schekochihin}, A.~A., {Cowley}, S.~C.,
  \& {Yousef}, T.~A. 2009, \mnras, 399, L146

\bibitem[{{Matthaeus} \& {Lamkin}(1986)}]{1986PhFl...29.2513M}
{Matthaeus}, W.~H., \& {Lamkin}, S.~L. 1986, Phys. Fluids, 29, 2513

\bibitem[{{Ng} et~al.(2008){Ng}, {Rosenberg}, {Germaschewski}, {Pouquet}, \&
  {Bhattacharjee}}]{2008ApJS..177..613N}
{Ng}, C.~S., {Rosenberg}, D., {Germaschewski}, K., {Pouquet}, A., \&
  {Bhattacharjee}, A. 2008, \apjs, 177, 613

\end{thebibliography}

\end{document}